\documentclass{PoS}
\usepackage[utf8]{inputenc}
\usepackage{amsmath}
\usepackage{listings}

\def\email#1{\texttt{#1}}

\def\f{\mathbf{f}}
\def\g{\mathbf{g}}

\def\z{\mathbf{z}}
\def\y{\mathbf{y}}

\definecolor{shaded}{RGB}{252,252,252}
\title{Analytic multi-loop results using finite fields and dataflow graphs with FiniteFlow}

\ShortTitle{Analytic multi-loop results using finite fields and dataflow graphs with FiniteFlow}

\author{\speaker{Tiziano Peraro}\\
        Physik-Institut, Universit\"at Z\"urich, Wintherturerstrasse 190, CH-8057 Z\"urich, Switzerland\\
        E-mail: \email{peraro@physik.uzh.ch}}

      \abstract{\textsc{FiniteFlow} is a public framework for defining
        and executing numerical algorithms over finite fields and
        reconstructing multivariate rational functions.  The framework
        allows to build complex algorithms by combining basic building
        blocks into graphical representations of the calculation,
        known as dataflow graphs.  It offers an easy-to-use
        \textsc{Mathematica} interface for implementing efficient
        custom algorithms without any low-level coding.  We report on
        some new features of \textsc{FiniteFlow} which have been
        published after its initial release, give some simple example
        of usage for common tasks and review recent cutting-edge
        applications to two-loop five-point scattering and the
        four-loop cusp anomalous dimension.}

\FullConference{14th International Symposium on Radiative Corrections (RADCOR2019)\\
9-13 September 2019\\
		Palais des Papes, Avignon, France}

\begin{document}

\section{Introduction}

State-of-the-art phenomenological and theoretical problems in
high-energy physics demand high-precision calculations for complex
processes.  These are characterized by either a high number of loops,
for higher precision, or many external legs and physical scales, for
processes involving more complicated combinations of external and
internal states.  Modern collider experiments at the LHC are sensitive
to these kinds of theoretical predictions, which therefore can be an
essential ingredient for validating the Standard Model in wider
regions of the phase-space and distinguishing possible signals of new
or interesting physics from the large Standard Model background.

In recent years, great effort has been put into the development of
efficient methods for complex multi-loop calculations.  A recent
breakthrough stemmed from the idea of reconstructing analytic results
from numerical evaluations over finite fields, where each arithmetic
operation is performed modulo a prime integer.  While some of these
ideas had already been used in other fields, such as computer algebra,
their introduction into high-energy
physics~\cite{vonManteuffel:2014ixa,Peraro:2016wsq} yielded several
impressive results (see e.g.\
ref.s~~\cite{vonManteuffel:2016xki,Badger:2018enw,Abreu:2018zmy,Lee:2019zop,Henn:2019rmi,vonManteuffel:2019wbj,Abreu:2019odu,vonManteuffel:2019gpr,Badger:2019djh})
which are beyond the capabilities of traditional computational tools.

A number of these
results~\cite{Badger:2017jhb,Badger:2018enw,Henn:2019rmi,Badger:2019djh,Hartanto:2019uvl}
have been obtained using \textsc{FiniteFlow}~\cite{Peraro:2019svx}, a
framework for defining and executing numerical algorithms over finite
fields and reconstructing multivariate rational functions.  In these
proceedings, we briefly review some of its main features and applications
and report on some new procedures and algorithms implemented and
published after its initial release.

\section{The FiniteFlow framework}

In this section, we give a brief summary of the most important features
of the \textsc{FiniteFlow} framework.  \textsc{FiniteFlow} consists of
three main components
\begin{itemize}
\item An efficient low-level implementation of a number of common
  basic algorithms (evaluation of functions, linear solvers and many
  more).
\item A framework for combining these basic algorithms, used as
  building blocks, into more complex ones, by means of computational
  graphs known as \emph{dataflow graphs}.  The basic algorithms are
  used as nodes of the graph and their output can be used as input by
  other nodes.
\item Functional reconstruction techniques yielding analytic
  expressions from numerical evaluations.
\end{itemize}
In \textsc{FiniteFlow} we use a simplified implementation of dataflow
graphs where \emph{nodes}, i.e.\ our basic algorithms, can have
one or more lists of numerical values, each represented by an
\emph{arrow}, as input, and exactly one list, i.e.\ one arrow, as
output.  A special node called \emph{input node} represents the input
variables of the numerical algorithm defined by the graph.  A node is
then identified as the \emph{output node}, whose output is identified
as the output of the graph itself.

These computational graphs can be easily defined by combining basic
algorithms into computational graphs, from computer algebra systems
and high-level programming languages (at the moment we provide a
\textsc{Mathematica} interface).  This allows to implement a wide
number of algorithms without any low-level coding or compilation step.

The \texttt{finiteflow} repository contains the source code of a
proof-of-concept implementation of this framework, as well as
instructions and tutorials for its installation and usage.  The
\texttt{finiteflow-mathtools} repository contains a number of
\textsc{Mathematica} packages and examples using \textsc{FiniteFlow}
for the solution of a number of extremely relevant problems in
high-energy physics.  Both repositories can be accessed at the URL
\texttt{{https://github.com/peraro}}.

A more comprehensive description of the \textsc{FiniteFlow} framework
and its implementation is given in~\cite{Peraro:2019svx}.

\section{New features and algorithms}

In this section, we report on some new features added to
\textsc{FiniteFlow} after its initial release.  These include new
algorithms which can be used as nodes of computational graphs as well
as additional procedures and options.

\subsection{Arbitrary rational expressions}
\label{sec:ratexpreval}

Most calculations start from some analytic input which can be cast in
the form of one or more lists of rational functions.  In
\textsc{FiniteFlow}, the preferred method for using these lists of
functions is embedding them in a node which evaluates them efficiently
using the so-called Horner's scheme (in the \textsc{Mathematica}
interface, this is done via the procedure \texttt{FFAlgRatFunEval}).
While this works extremely well in most cases, there are a few
situations where it is not ideal.  Indeed, in order to use the
Horner scheme for a rational function, one must first write it as a
ratio of two polynomials (not necessarily GCD-simplified) and expand
its numerator and denominator.  In some cases, however, performing
such operations may be difficult, due to the complexity of the
analytic input, or may yield an expression which is considerably more
complicated than the original one, thus negatively affecting
performance.

We stress that, in realistic problems, one can almost always sidestep
these issues by implementing larger portions of an algorithm using
\textsc{FiniteFlow}, rather than analytically (see e.g.\
section~\ref{sec:examples} and the examples in the public
repositories).  However, for the few cases where this is impossible or
inconvenient, we have implemented an algorithm which allows to
evaluate arbitrary \emph{rational expressions} in a list of variables
$\z$, without the need of preliminarily casting them into any special
form.

More in detail, rational expressions are ``compiled'' into
\emph{bytecode}, i.e.\ an array of bytes representing a sequence of
rational operations needed to evaluate the functions numerically (this
is all done at runtime and no actual compilation to machine code is
performed).  In the \textsc{Mathematica} interface, this is done via
the procedure \texttt{FFAlgRatExprEval}.  The bytecode is currently
generated directly in \textsc{Mathematica} by recursively inspecting
the list of expressions and all their subexpressions.  The evaluation
of the function in \textsc{C++} is done via a simple \emph{virtual
  machine}, i.e.\ a procedure which reads the bytecode in a loop and
executes the corresponding instructions.  At the moment no
optimization (such as reuse of common subexpressions) is performed,
although these can be implemented directly in \textsc{Mathematica} by
combining several nodes of this type.

As we mentioned, this algorithm can be extremely useful when dealing
with functions which are too hard to manipulate analytically or when
the functions are significantly simpler in a form which is
\emph{not} the ratio of two expanded polynomials.  It may also be more
efficient for very sparse functions.

As an example, we applied this to a complicated function in three
variables, which is about 1.5~GB in plain text form.  The definition
of the node, including the generation of the bytecode in
\textsc{Mathematica}, takes about 10 minutes on a modern machine.  The
numerical evaluation of the list of functions then takes about 0.5
seconds per point.

\subsection{Rational functions with parametric coefficients}
\label{sec:rati-funct-with}

Consider a generic rational function in the variables
$\z=\{z_1,z_2,\ldots\}$ with the form
\begin{equation}
  \label{eq:ratfun}
  f(\z) = \frac{\sum_\alpha\, n_\alpha\, \z^\alpha}{\sum_\alpha\, d_\alpha\, \z^\alpha}.
\end{equation}
As already mentioned, \textsc{FiniteFlow} provides routines for
evaluating functions from their analytic expressions, cast as in the
previous equation where $n_\alpha$ and $d_\alpha$ are rational
numbers, using Horner's method.  However, there are cases where the
coefficients $n_\alpha$ and $d_\alpha$ are not known rational numbers
but they can be computed numerically as the output of a node in a
computational graph.  We have therefore added an alternative to the
standard evaluation algorithms, called
\texttt{FFAlgRatFunEvalFromCoeffs} in the \textsc{Mathematica}
interface.  This node takes two lists as input, i.e.\ a list of values
for the variables $\z$ and a list of values for the coefficients
$\{n_\alpha,d_\alpha\}$.

\subsection{Partial reconstruction of subgraphs}
\label{sec:part-reconstr-subgr}

\textsc{FiniteFlow} can define some type of nodes in a graph called
\emph{subgraph} nodes, that are nodes which evaluate another graph (a
subgraph) a number of times in order to produce their output (more
detailes can be found in~\cite{Peraro:2019svx}).  We added a new type
of subgraph node, which performs a partial functional reconstruction
of the output of another graph.

Let $G=G(\z,\y)$ be a graph representing a rational function in two
sets of variables $\z$ and $\y$.  The \emph{subgraph reconstruction}
algorithm is a node of another graph, which takes a list of values for
$\y$ as input, evaluates $G$ a number of times, and reconstructs its
output as a function of $\z$ only, as on the right-hand side of
eq.~\eqref{eq:ratfun}.  It returns a list of numerical coefficients
$n_\alpha$ and $d_\alpha$ as functions of the numerical values of $\y$
which are the input of the node.  Hence, the (numerical) coefficients
$\{n_\alpha,d_\alpha\}$ may subsequently be used as inputs for other
nodes of the graph.

An example where this might be useful is when $\z$ are loop
propagators and $\y$ are kinematic variables.  One may want to
reconstruct the dependence of the integrand on the propagators $\z$
and only compute the kinematic coefficients numerically so that they
can be used in other steps of the calculation (such as Integration By
Parts) \emph{before} the full analytic reconstruction.  In other
words, this can be used, in some cases, as an implementation of
\emph{integrand reduction}~\cite{Ossola:2006us,Giele:2008ve,Mastrolia:2011pr,Badger:2012dp,Zhang:2012ce,Mastrolia:2012an}
which employs the functional reconstruction algorithms of
\textsc{FiniteFlow} rather than evaluations on cuts and system
solving.

\subsection{Sparse output for sparse solvers}
\label{sec:sparse-output-sparse}

We added the option to return a \emph{sparse output} from a sparse
linear solver node.  By default, linear solvers, in the general case
where there are fewer independent equations than unknowns, return a
dense matrix of coefficients which convert the linearly dependent
unknowns into linearly independent ones (e.g.\ a matrix converting a
list of integrals to a linear combination of master integrals, in the
case of Integration By Parts).  There are cases where one may wish to
return a sparse representation of this matrix, i.e.\ only the
non-vanishing coefficients of the solution.  This effect could already
be achieved by combining a linear solver with the \texttt{NonZero}
algorithm (see ref.~\cite{Peraro:2019svx}) but it is now possible to
return such a representation from a sparse solver node directly, as an
option (this is done using \texttt{FFSolverSparseOutput} from the
\textsc{Mathematica} interface).

\subsection{Parallel evaluation of user-defined lists of points}
\label{sec:parall-eval-user}

We added a procedure for evaluating a graph for a user-specified list of
sample points.  The procedure is called \texttt{FFGraphEvaluateMany}
in the \textsc{Mathematica} interface.  By default, it evaluates the
points in parallel, using multi-threading.  The output should be
equivalent to calling the \texttt{FFGraphEvaluate} procedure (which
has been present since the first release of \textsc{FiniteFlow}) on
each individual point, except that the evaluations can be parallelized
with the new procedure.  This can be useful any time one needs to
evaluate a graph for a large number of sample points, e.g.\ for using
custom reconstruction algorithms or finding linear relations between
smaller subsets of the full output.

As an example, consider a process whose result might be too hard or
inconvenient to reconstruct analytically.  If, for the purposes of a
particular study, it is sufficient to evaluate an amplitude
numerically for a number of phase-space points which is not too large,
one may proceed by building a graph and reconstructing its numerical
output for these points.\footnote{Here we refer to the part of the
  calculation which can be performed via rational operations and thus
  implemented over finite fields, for instance the reduction of an
  amplitude to master integrals.  If the master integrals themselves
  are not known, we understand that one also needs a method to compute
  those numerically, e.g.\ using sector
  decomposition~\cite{Binoth:2000ps}.}  More in detail, for each
phase-space point, one can consider a numerical rational approximation
of the independent kinematic invariants which is correct up to a given
accuracy.  Then we evaluate the graph at these points over several
prime fields, which can be done in parallel using
\texttt{FFGraphEvaluateMany}, and use them to reconstruct the output
over the rational field.  From the \textsc{Mathematica} interface, one
can use the \texttt{FFRatMod} to convert a rational number to an
integer over a finite field.  The rational reconstruction can be done
(see e.g.~\cite{Peraro:2016wsq,ManuelThesis} for a pedagogical description)
by combining the Chinese remainder theorem (using e.g.\ the builtin
\texttt{ChineseRemainder} routine of \textsc{Mathematica}) and Wang's
reconstruction (using e.g.\ the \texttt{FFRatRec} procedure of
\textsc{FiniteFlow}).  A similar approach can obviously be implemented
directly in the \textsc{C++} interface as well.

While this is not as
performant as a floating-point calculation, it is free of numerical
instabilities and provides an exact result (up to the accuracy of the
rationalization one started with).  Moreover, this strategy allows to
exploit the computational tools and high-level interface of the
\textsc{FiniteFlow} framework, which can drastically reduce
development time compared to a custom implementation of the same
algorithm in a low-level language.

\section{Simple examples}
\label{sec:examples}

In this section, we illustrate some examples of usage of
\textsc{FiniteFlow} for simple but common tasks.  Some prior
familiarity with the \textsc{FiniteFlow} code might be beneficial to
the reader.  For an introduction to the usage of the code, we refer to
the \textsc{Mathematica} tutorial in the main repository and to the
examples in the \texttt{finiteflow-mathtools} repository.

\subsection{Linear substitutions}
\label{sec:linsub}

Linear substitutions are among the most ubiquitous algebraic
operations in multi-loop calculations.  Performing such substitutions
analytically sometimes results in an explosion of the algebraic
complexity of the expressions, before the final GCD simplification of
the result.  These substitutions can, however, be reinterpreted as
matrix multiplications, which in turn can be easily performed using
\textsc{FiniteFlow}.

Let us consider a list of quantities $A_j$, such as a list of
amplitudes or form factors for a process, which can be written as
linear combinations of some objects $f_k$, representing e.g.\ Feynman
integrals or other special functions,
\begin{equation}
  \label{eq:Aj}
  A_j = \sum_j a_{jk}\ f_k,
\end{equation}
where the coefficients $a_{jk}=a_{jk}(\z)$ are \emph{rational
  functions} of some variables $\z=\{z_1,z_2,\ldots\}$.  Let us
consider also a set of linear relations between the $f_j$ and
another set of objects $g_k$,
\begin{equation}
  \label{eq:funred}
  f_j = \sum_k c_{jk}\, g_k,
\end{equation}
where the coefficients $c_{jk}=c_{jk}(\z)$ are also \emph{rational
  functions} of $z$.  The previous equation may represent e.g.\ the
reduction of a set of Feynman integrals $f_j$ to master integrals
$g_k$, or the explicit analytic solution for the integrals $f_j$ in
terms of a list of special functions $g_k$, or any other relation with
a similar form.

We thus wish to substitute eq.~\eqref{eq:funred} into
eq.~\eqref{eq:Aj} in order to express $A_j$ in terms of $g_k$.  As
already mentioned, we can
use the matrix multiplication routines of \textsc{FiniteFlow} to
perform such substitution numerically and reconstruct the final result
analytically.  In other words, we want to reconstruct the matrix of
coefficients $x_{jk}=x_{jk}(\z)$ such that
\begin{equation}
  \label{eq:3}
  A_j = \sum_{jk} x_{jk} \, g_k
\end{equation}
which are given by the matrix multiplication
\begin{equation}
  \label{eq:1}
  x_{jk} = \sum_l a_{jl}\, c_{lk}.
\end{equation}

In fig.~\ref{sec:linsub} we show a possible implementation of this in
\textsc{FiniteFlow}, assuming that the coefficients $a_{jk}$ and
$c_{jk}$ above are known analytically.  A similar approach can, of
course, be used when the coefficients are known numerically and
computed by other nodes of the graph.
\begin{figure}[t]
  \centering
  \begin{lstlisting}
  (* Input variables *)
  zs = {$z_1, z_2, \ldots $};

  (* Analytic input matrices and their dimension  *)
  ajkmat = {{$a_{11}(\z),a_{12}(\z),\ldots$},$\ldots$};
  cjkmat = {{$c_{11}(\z),c_{12}(\z),\ldots$},$\ldots$};
  {n1,n2} = Dimensions[ajkmat];
  {m1,m2} = Dimensions[cjkmat]; (* must satisfy n2 == m1 *)

  (* Implement multiplication in FiniteFlow *)
  FFNewGraph[graph,input,zs];
  FFAlgRatFunEval[graph,ajk,{input},zs,Join@@ajkmat];
  FFAlgRatFunEval[graph,cjk,{input},zs,Join@@cjkmat];
  FFAlgMatMul[graph,xjk,{ajk,cjk},n1,n2,m2];
  FFGraphOutput[graph,xjk];

  (* Reconstruct and reshape the result *)
  result = FFReconstructFunction[graph,zs];
  result = ArrayReshape[result,{n1,m2}];
\end{lstlisting}\label{fig:linsub}
  \caption{An implementation of linear substitutions in
    \textsc{FiniteFlow}.}
\end{figure}

\subsection{Changes of variables}

Changes of variables are another common operation.  This is trivially
implemented in \textsc{FiniteFlow} by concatenating nodes that
evaluate rational functions.

More explicitly, suppose we have a list of functions
\begin{equation}
  \label{eq:2}
 \f(\y)=\{f_1(\y),f_2(\y),\ldots\}
\end{equation}
of the variables $\y$ and that we want to perform a change of
variables from $\y$ to $\z$
\begin{equation}
  \label{eq:4}
  \y = \y(\z) = \{y_1(\z),y_2(\z),\ldots\}.
\end{equation}
In other words, we want to compute a list of functions $\g=\g(\z)$
such that
\begin{equation}
  \label{eq:5}
  \g(\z) = \{g_1(\z),g_2(\z),\ldots\} = \f(\y(\z)).
\end{equation}
Assuming the initial list of functions $\f$ and the change of
variables $\y=\y(\z)$ are known analytically, one can implement this,
as in fig.~\ref{fig:chvar}, simply by creating a node which evaluates
$\y=\y(\z)$ and use it as input for a node which evaluates
$\f=\f(\y)$.
\begin{figure}[t]
  \centering
  \begin{lstlisting}
  zs = {$z_1, z_2, \ldots $};
  FFNewGraph[graph,input,zs];
  FFAlgRatFunEval[graph,y,{input},zs,{$y_1(\z),y_2(\z),\ldots$}];
  FFAlgRatFunEval[graph,f,{y},{$y_1,y_2,\ldots$},{$f_1(\y),f_2(\y),\ldots$}];
  FFGraphOutput[graph,f];
  result = FFReconstructFunction[graph,zs];
\end{lstlisting}
  \caption{A change of variables in \textsc{FiniteFlow}.}
  \label{fig:chvar}
\end{figure}

\subsection{GCD simplification}
All the analytic results reconstructed by \textsc{FiniteFlow} are
automatically GCD-simplified, i.e.\ numerator and denominator have
minimal degrees and no common factor, as a feature of the
functional reconstruction algorithm.  Hence, one can perform a
GCD-simplification simply by asking \textsc{FiniteFlow} to reconstruct
the analytic expression of the corresponding rational function.

If one has an analytic expression which is sufficiently simple or such
that the number of operations required for GCD simplification is not
too large, then the traditional simplification procedures of computer
algebra systems (such as \textsc{Mathematica}'s \texttt{Together}) are
likely to be more efficient than a full functional reconstruction.
However for very complicated expressions, where algebraic
manipulations are very expensive, one may consider building a
numerical algorithm evaluating that expression and reconstruct its
simplified result out of numerical evaluations with
\textsc{FiniteFlow}.

In order to build such an algorithm, one must combine the builtin
basic algorithms of \textsc{FiniteFlow} in a suitable computational
graph.  For this purpose, the examples above and the ones in the
public repositories might be useful as an inspiration.  The
\texttt{TakeAndAdd} algorithm of \textsc{FiniteFlow} can also be
useful when the expression is a sum of simple functions.

A simple (albeit not always optimal) strategy that can always be
applied consists in using the full expression to be simplified,
together with the new algorithm described in
section~\ref{sec:ratexpreval}, to quickly and effectively build the
required numerical algorithm.  This can be immediately followed by a
call to the reconstruction procedures.  For the convenience of the
users, we already provide a wrapper routine, called
\texttt{FFTogether}, which does this automatically for any expression
and can simply be used with
\begin{equation*}
  \mathtt{FFTogether[expression]}
\end{equation*}
where \texttt{expression} is a rational function (or a list, in which
case \texttt{FFTogether} is applied to each element).  The output is a
collected and GCD-simplified form of the input expression, i.e.\
roughly equivalent to \textsc{Mathematica}'s builtin \texttt{Together}
function.

\section{Cutting-edge applications}

In this section, we briefly summarize some cutting edge applications of
the \textsc{FiniteFlow} framework.  These show the effectiveness of
this framework as well as its versatility since it has been applied
to a variety of methods and techniques.

\subsection{Two-loop five-point scattering}

\textsc{FiniteFlow}, before and after its public release, has been
used in a number of published results for two-loop five-point
scattering.  These processes are extremely complicated, not only
because of the number of diagrams and the size of the systems of
Integration By Parts identities to be solved, but also for the
presence of at least five independent invariants.  In the following,
we briefly summarize the usage of \textsc{FiniteFlow} in the context
of these calculations, while we refer the reader to the corresponding
papers for further details.

In~\cite{Badger:2017jhb} the first numerical results for a full set of
two-loop five-gluon helicity amplitudes have been published.  We used
an implementation of integrand reduction in \textsc{FiniteFlow} and
sector decomposition to compute the integrals.

In~\cite{Badger:2018enw} we reconstructed the first analytic results
for the single-minus helicity configuration.  We used a custom
\textsc{C++} implementation of unitarity cuts and linear fits based
on~\cite{Peraro:2016wsq} (the implementation of unitarity cuts is not
public since it is no longer actively developed).  Using
\textsc{FiniteFlow}, we used this building block to implement a
computational graph containing a complete set of unitarity cuts. This
graph, for any phase-space point, reconstructs a complete integrand
representation.  This was then multiplied by the numerical output of
the solution of Integration-By-Parts, computed with the sparse linear
solver of \textsc{FiniteFlow}.  Finally, the result has been cast in
terms of known pentagon functions~\cite{Gehrmann:2018yef} and the
known divergence structure has been subtracted.  For each numerical
phase-space point, we performed a full reconstruction with respect to
the dimensional regulator $\epsilon$ and used it to extract the finite
part of its Laurent expansion around $\epsilon=0$.  This can be easily
done using the tools provided by \textsc{FiniteFlow}, directly from
the \textsc{Mathematica} interface.  Hence, the finite part has been
reconstructed analytically from these numerical evaluations, as a
function of kinematic invariants.  More details on the procedure can
be found in~\cite{Badger:2018enw,Peraro:2019svx} and explicit examples
of similar calculations are available in the public repositories.

In~\cite{Badger:2019djh} we computed analytically the two-loop
five-gluon all-plus helicity amplitude, including for the first time
the full-colour dependence.  The main difference with respect to the
previous results was the availability of an analytic integrand
representation~\cite{Badger:2015lda}. We used \textsc{FiniteFlow} to
reduce the linear combinations of integrals contributing to the
amplitude to master integrals.  Although this required rank five
reductions (and rank six for one-loop squared diagrams), since the
results were relatively simple, these were reconstructed analytically
in a few hours using a modern computing node with 32 threads.  From
this, assembling the full amplitude, summed over the 120 permutations
of the integrand in~\cite{Badger:2015lda}, and reconstructing it in
terms of iterated integrals (which we derived in the physical
$s_{12}$-scattering region) took only a few minutes on a modern
laptop.

In~\cite{Hartanto:2019uvl} we computed numerically a complete set of
two-loop helicity amplitudes involving four-partons and a $W$ boson.
The setup was similar to the one in~\cite{Badger:2018enw}, with some
key differences.  The integrand was generated by applying integrand
reduction to Feynman diagrams, rather than using unitarity cuts.  We
reconstructed numerically the coefficients of the master integrals.
The master integrals which are unknown have been computed using sector
decomposition.  In particular, we chose as masters some combinations
of integrals which have a simpler divergence structure, thus improving
performance and reliability of the sector decomposition method.

\textsc{FiniteFlow} has also been used in a number of unpublished
multi-leg and multi-scale calculations and tests, as well as for
simplifying complicated analytic expressions (see e.g.\
ref.~\cite{Chawdhry:2019bji}).

\subsection{Four-loop cusp anomalous dimension}

\begin{figure}[t]
  \centering
\includegraphics[width=0.21 \textwidth]{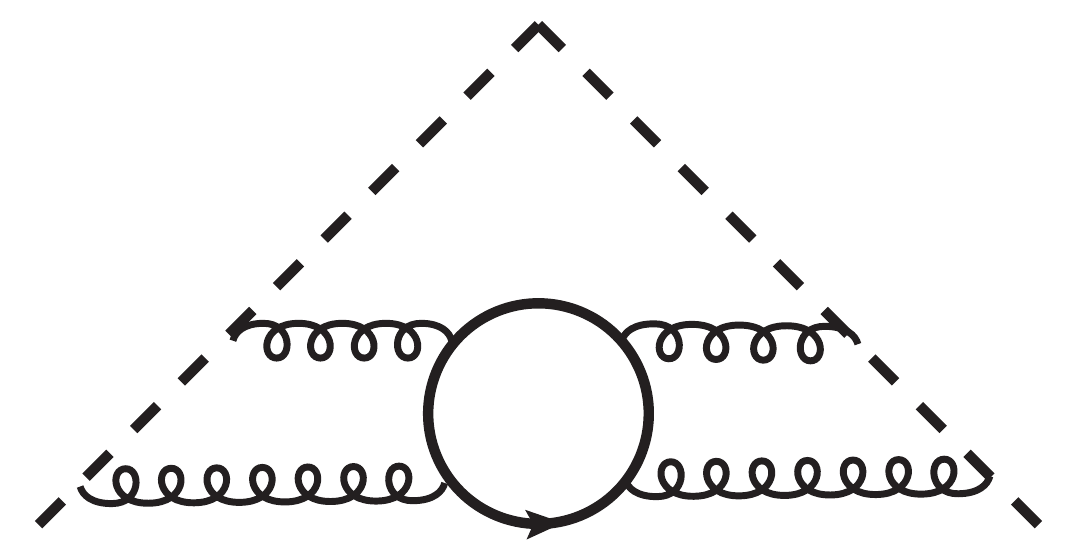}%
\includegraphics[width=0.21 \textwidth]{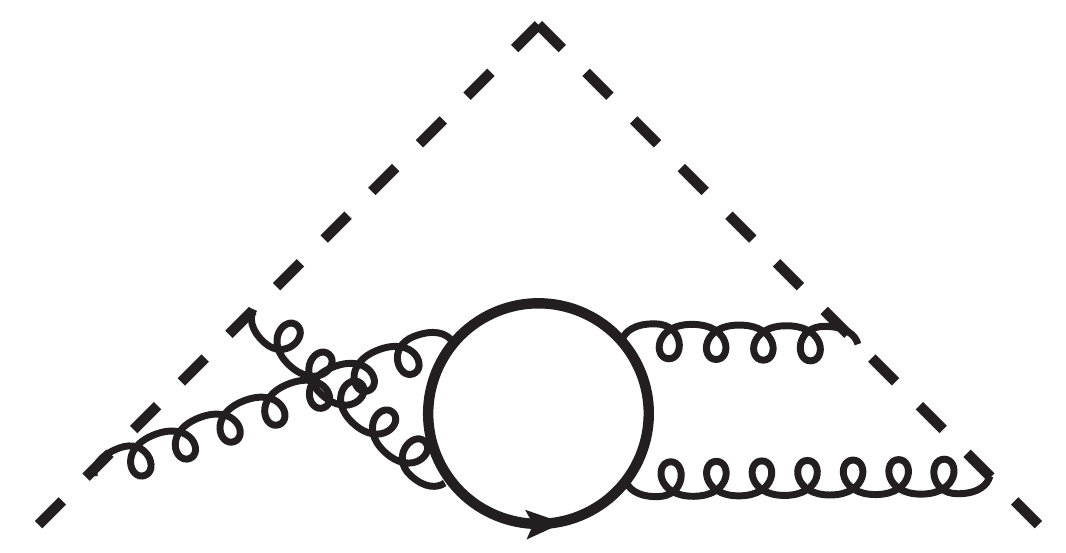}%
\includegraphics[width=0.21 \textwidth]{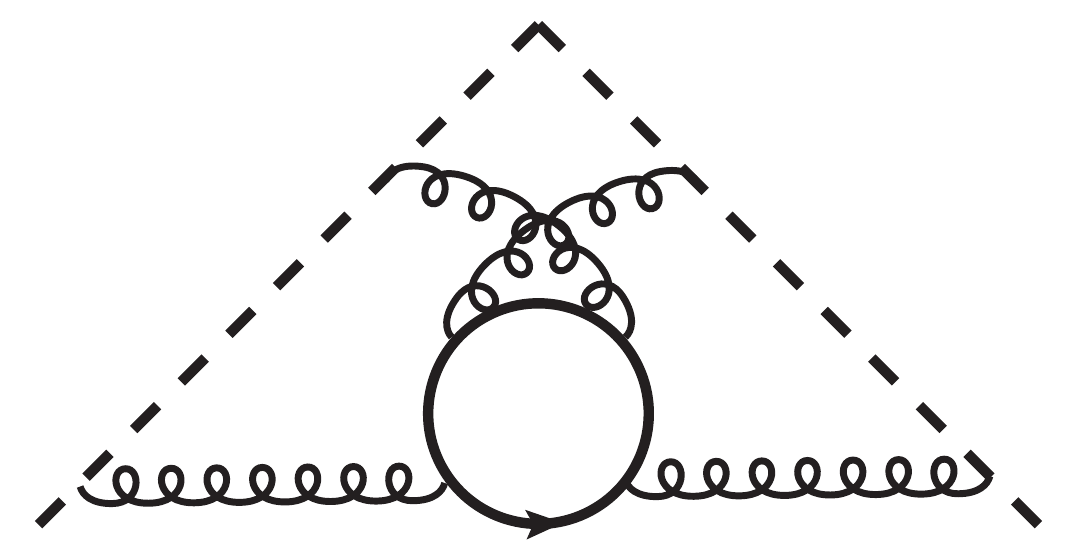}%
\\
\includegraphics[width=0.21 \textwidth]{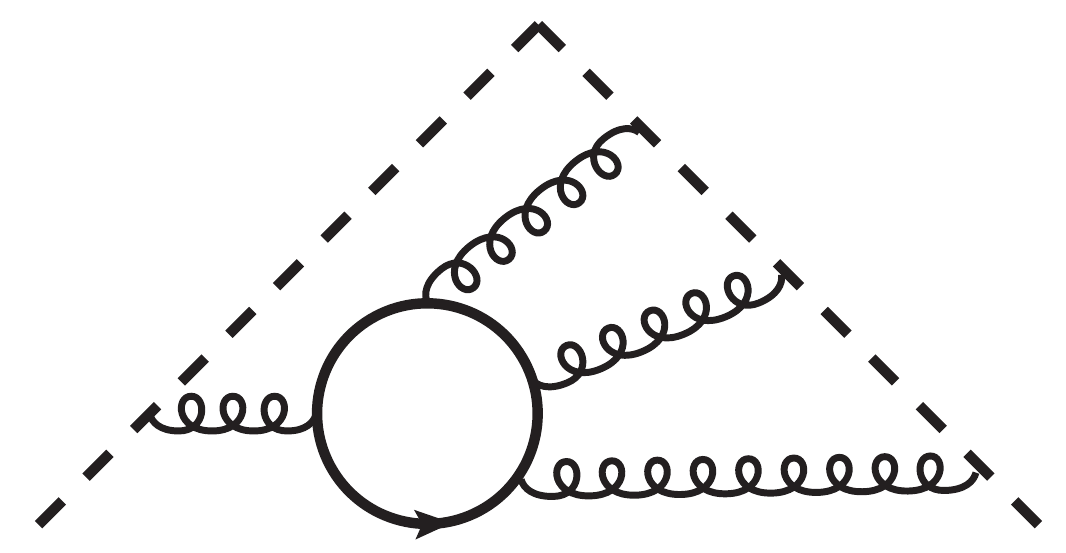}%
\includegraphics[width=0.21 \textwidth]{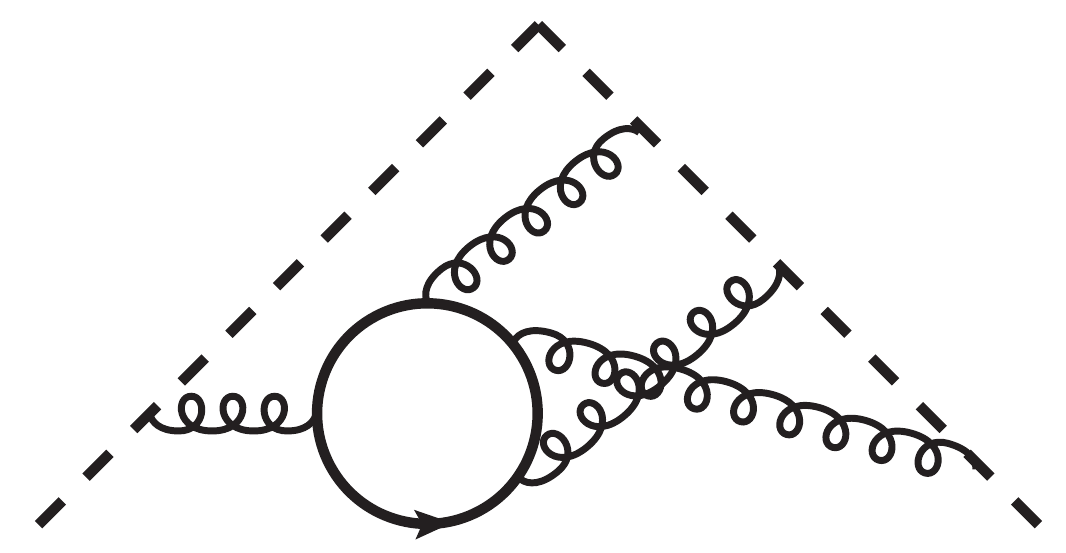}%
\includegraphics[width=0.21 \textwidth]{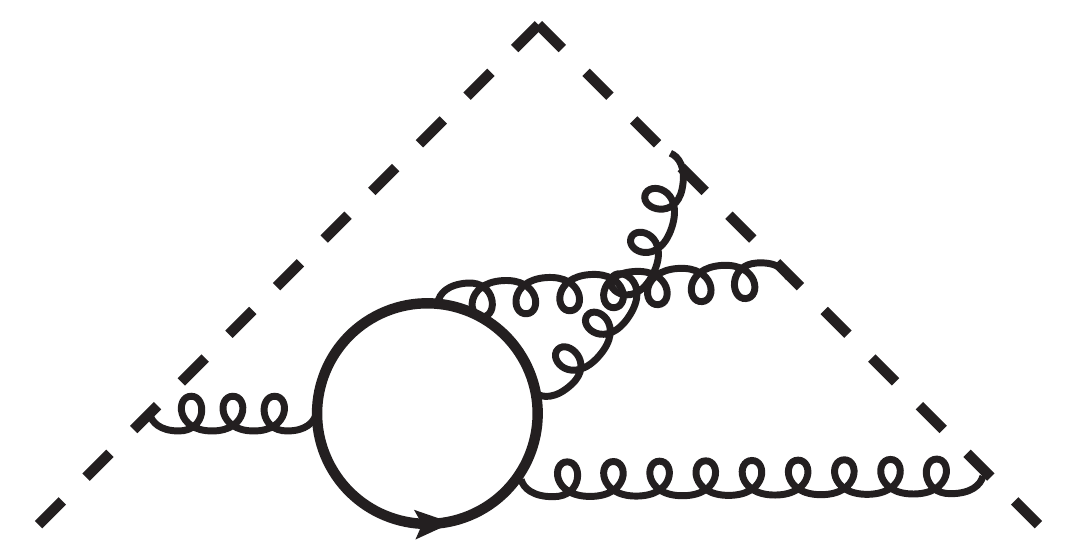}%
  \caption{Fermion-loop diagrams contributing to the quartic Casimir
    part of the four-loop light-like cusp anomalous dimension.}
  \label{fig:4loopcusp}
\end{figure}

The light-like cusp anomalous dimension is a universal quantity in
QCD.  It determines infrared divergences of scattering amplitudes and
also appears in the resummation of soft and collinear Sudakov double
logarithms.  The matter-dependent four-loop contributions coming from
both scalar and fermion loops have been computed analytically
in~\cite{Henn:2019rmi} (the fermionic contributions have also been
independently computed in~\cite{Lee:2019zop}).  The quartic Casimir
terms of the cusp anomalous dimension have been computed from the
double pole of a form factor, defined by a composite operator inserted
into on-shell scalar states (the other contributions have been derived
using a conjecture in a parallel publication~\cite{Bruser:2019auj}).
The main computational bottleneck of the calculation was the reduction
of the relevant four-loop diagrams to master integrals.  The reduction
was performed using the sparse solver of \textsc{FiniteFlow}.  Rather
than producing large and complex reduction tables, we reconstructed
analytically the reduction of each individual diagram to a set of
master integrals, which have been chosen for their analytic
properties, namely uniform and maximal transcendental weight and
simpler pole structure (see ref.~\cite{Henn:2019rmi} for more
details).  We also used \textsc{FiniteFlow} for producing some of the
differential equations satisfied by off-shell generalizations of these
integrals, which have been used for their analytic calculation.

Unlike the five-point case, where the complexity is mostly a
consequence of the large number of scales, in this application, there
is only one scale and the complexity is due to the huge size of the
systems of equations to be solved for the reduction to master
integrals.  This shows that \textsc{FiniteFlow} can be effectively
used for a wide and diverse range of complex multi-loop calculations.

\section{Conclusions}
\label{sec:conclusions}

\textsc{FiniteFlow} is a powerful framework for performing complex
algebraic calculations using modern techniques.  It allows to define
complicated numerical algorithms over finite fields via a high-level
interface based on dataflow graphs and reconstruct full analytic
results using functional reconstruction techniques.  A number of
recent
applications~\cite{Badger:2017jhb,Badger:2018enw,Henn:2019rmi,Badger:2019djh,Hartanto:2019uvl}
showed the capability of this framework of pushing the
state-of-the-art for theoretical predictions in high-energy physics.

In these proceedings, we reported on new features published after its
initial release, we provided simple examples of its usage and we
reviewed some of its most relevant applications.

\section*{Acknowledgements}
I thank Simon Badger and Pierpaolo Mastrolia for useful comments.  I
am also grateful to Simon Badger, Christian Brønnum-Hansen, Christoph
Dlapa, Heribertus Bayu Hartanto, William Torres Bobadilla, and Simone
Zoia for proving valuable feedback on the \textsc{FiniteFlow} program.
This work has received funding from the European Union’s Horizon 2020
research and innovation programme under the Marie Skłodowska-Curie
grant agreement 746223.


\begin{thebibliography}{99}
\bibitem{vonManteuffel:2014ixa}
  A.~von Manteuffel and R.~M.~Schabinger,
  Phys.\ Lett.\ B {\bf 744} (2015) 101
  doi:10.1016/j.physletb.2015.03.029
  [arXiv:1406.4513 [hep-ph]].


\bibitem{Peraro:2016wsq}
  T.~Peraro,
  JHEP {\bf 1612} (2016) 030
  doi:10.1007/JHEP12(2016)030
  [arXiv:1608.01902 [hep-ph]].


\bibitem{vonManteuffel:2016xki}
  A.~von Manteuffel and R.~M.~Schabinger,
  Phys.\ Rev.\ D {\bf 95} (2017) no.3,  034030
  doi:10.1103/PhysRevD.95.034030
  [arXiv:1611.00795 [hep-ph]].


\bibitem{Badger:2018enw}
  S.~Badger, C.~Brønnum-Hansen, H.~B.~Hartanto and T.~Peraro,
  JHEP {\bf 1901} (2019) 186
  doi:10.1007/JHEP01(2019)186
  [arXiv:1811.11699 [hep-ph]].


\bibitem{Abreu:2018zmy}
  S.~Abreu, J.~Dormans, F.~Febres Cordero, H.~Ita and B.~Page,
  Phys.\ Rev.\ Lett.\  {\bf 122} (2019) no.8,  082002
  doi:10.1103/PhysRevLett.122.082002
  [arXiv:1812.04586 [hep-ph]].


\bibitem{Lee:2019zop}
  R.~N.~Lee, A.~V.~Smirnov, V.~A.~Smirnov and M.~Steinhauser,
  JHEP {\bf 1902} (2019) 172
  doi:10.1007/JHEP02(2019)172
  [arXiv:1901.02898 [hep-ph]].


\bibitem{Henn:2019rmi}
  J.~M.~Henn, T.~Peraro, M.~Stahlhofen and P.~Wasser,
  Phys.\ Rev.\ Lett.\  {\bf 122} (2019) no.20,  201602
  doi:10.1103/PhysRevLett.122.201602
  [arXiv:1901.03693 [hep-ph]].


\bibitem{vonManteuffel:2019wbj}
  A.~von Manteuffel and R.~M.~Schabinger,
  Phys.\ Rev.\ D {\bf 99} (2019) no.9,  094014
  doi:10.1103/PhysRevD.99.094014
  [arXiv:1902.08208 [hep-ph]].


\bibitem{Abreu:2019odu}
  S.~Abreu, J.~Dormans, F.~Febres Cordero, H.~Ita, B.~Page and V.~Sotnikov,
  JHEP {\bf 1905} (2019) 084
  doi:10.1007/JHEP05(2019)084
  [arXiv:1904.00945 [hep-ph]].


\bibitem{vonManteuffel:2019gpr}
  A.~von Manteuffel and R.~M.~Schabinger,
  JHEP {\bf 1905} (2019) 073
  doi:10.1007/JHEP05(2019)073
  [arXiv:1903.06171 [hep-ph]].


\bibitem{Badger:2019djh}
  S.~Badger {\it et al.},
  Phys.\ Rev.\ Lett.\  {\bf 123} (2019) no.7,  071601
  doi:10.1103/PhysRevLett.123.071601
  [arXiv:1905.03733 [hep-ph]].


\bibitem{Badger:2017jhb}
  S.~Badger, C.~Brønnum-Hansen, H.~B.~Hartanto and T.~Peraro,
  Phys.\ Rev.\ Lett.\  {\bf 120} (2018) no.9,  092001
  doi:10.1103/PhysRevLett.120.092001
  [arXiv:1712.02229 [hep-ph]].


\bibitem{Hartanto:2019uvl}
  H.~B.~Hartanto, S.~Badger, C.~Brønnum-Hansen and T.~Peraro,
  JHEP {\bf 1909} (2019) 119
  doi:10.1007/JHEP09(2019)119
  [arXiv:1906.11862 [hep-ph]].


\bibitem{Peraro:2019svx}
  T.~Peraro,
  JHEP {\bf 1907} (2019) 031
  doi:10.1007/JHEP07(2019)031
  [arXiv:1905.08019 [hep-ph]].


\bibitem{Ossola:2006us}
  G.~Ossola, C.~G.~Papadopoulos and R.~Pittau,
  Nucl.\ Phys.\ B {\bf 763} (2007) 147
  doi:10.1016/j.nuclphysb.2006.11.012
  [hep-ph/0609007].


\bibitem{Giele:2008ve}
  W.~T.~Giele, Z.~Kunszt and K.~Melnikov,
  JHEP {\bf 0804} (2008) 049
  doi:10.1088/1126-6708/2008/04/049
  [arXiv:0801.2237 [hep-ph]].


\bibitem{Mastrolia:2011pr}
  P.~Mastrolia and G.~Ossola,
  JHEP {\bf 1111} (2011) 014
  doi:10.1007/JHEP11(2011)014
  [arXiv:1107.6041 [hep-ph]].


\bibitem{Badger:2012dp}
  S.~Badger, H.~Frellesvig and Y.~Zhang,
  JHEP {\bf 1204} (2012) 055
  doi:10.1007/JHEP04(2012)055
  [arXiv:1202.2019 [hep-ph]].


\bibitem{Zhang:2012ce}
  Y.~Zhang,
  JHEP {\bf 1209} (2012) 042
  doi:10.1007/JHEP09(2012)042
  [arXiv:1205.5707 [hep-ph]].


\bibitem{Mastrolia:2012an}
  P.~Mastrolia, E.~Mirabella, G.~Ossola and T.~Peraro,
  Phys.\ Lett.\ B {\bf 718} (2012) 173
  doi:10.1016/j.physletb.2012.09.053
  [arXiv:1205.7087 [hep-ph]].


\bibitem{Binoth:2000ps}
  T.~Binoth and G.~Heinrich,
  Nucl.\ Phys.\ B {\bf 585} (2000) 741
  doi:10.1016/S0550-3213(00)00429-6
  [hep-ph/0004013].


\bibitem{ManuelThesis}
  M.~Accettulli~Huber,
  ``The natural structure of scattering amplitudes,''
 Master thesis.


\bibitem{Gehrmann:2018yef}
  T.~Gehrmann, J.~M.~Henn and N.~A.~Lo Presti,
  JHEP {\bf 1810} (2018) 103
  doi:10.1007/JHEP10(2018)103
  [arXiv:1807.09812 [hep-ph]].


\bibitem{Badger:2015lda}
  S.~Badger, G.~Mogull, A.~Ochirov and D.~O'Connell,
  JHEP {\bf 1510} (2015) 064
  doi:10.1007/JHEP10(2015)064
  [arXiv:1507.08797 [hep-ph]].


\bibitem{Chawdhry:2019bji}
  H.~A.~Chawdhry, M.~L.~Czakon, A.~Mitov and R.~Poncelet,
  arXiv:1911.00479 [hep-ph].


\bibitem{Bruser:2019auj}
  R.~Brüser, A.~Grozin, J.~M.~Henn and M.~Stahlhofen,
  JHEP {\bf 1905} (2019) 186
  doi:10.1007/JHEP05(2019)186
  [arXiv:1902.05076 [hep-ph]].

\end{thebibliography}
\end{document}